\documentclass[twocolumn,twoside,slac_two]{revtex4}
\usepackage{amsfonts,amsmath,amssymb}
\usepackage{graphicx}
\usepackage{fancyhdr}
\newcounter{saveeqn}
\newcommand{\alphaeqn}{\setcounter{saveeqn}{\value{equation}}%
\stepcounter{saveeqn}\setcounter{equation}{0}%
\renewcommand{\theequation}{\mbox{\arabic{saveeqn}\alph{equation}}}}
\newcommand{\reseteqn}{\setcounter{equation}{\value{saveeqn}}%
\renewcommand{\theequation}{\arabic{equation}}}

\begin{document}

\title{Could Leptons, Quarks or both be  Highly Relativistic Bound States of  Minimally Interacting Fermion and Scalar?}

%

\author{G. B. Mainland }
\affiliation{Department of Physics, The Ohio  State University at Newark, Newark, OH, 43055-1797, USA}

\begin{abstract}
The possibility that leptons, quarks or both might be highly relativistic bound states of  a spin-0 and spin-1/2 constituent bound by minimal electrodynamics is discussed. Typically,  strongly bound solutions of the Bethe-Salpeter equation exist only when the coupling constant is on the order of or greater than unity. For the bound-state system discussed here, there exist two classes of boundary conditions that could yield strongly bound solutions with coupling constants on the order of the electromagnetic fine structure constant $\alpha$. In both classes the bound state must have spin one half, thus providing a possible explanation for the absence of higher-spin  leptons and quarks.
\end{abstract}

\maketitle

\thispagestyle{fancy}

\section{Motivation for Composite Models of Leptons and Quarks }

The  motivation for studying composite models of leptons and quarks is almost entirely circumstantial. 
 Three times in the past 150 years  particles that were - or are currently - thought to be elementary could be organized into families:  (1)  In the 1860's Meyer and Mendelev arranged the atoms into the periodic table that consists of 18 families.  The existence of families was ultimately explained by the realization that atoms are composite.   (2) Almost 100 years later two of the then-fundamental particles, the neutron and proton - as well as many other baryons and mesons - were placed into  SU(3) multiplets with the ultimate result that the neutron and proton were no longer viewed as being fundamental but instead were comprised of quarks \cite{Gell-Mann:64,Zweig:64}.  (3) By the mid 1970's many physicists realized that the charged leptons, the neutrinos, the positively charged quarks and the negatively charged quarks constitute four families. Today the three charged families are known to have three members each, and the neutrino family  almost certainly has three members as well \cite{Amsler:08}.  Because the existence  of families of ``fundamental'' particles has twice  been explained by the realization that the particles are actually composite, although speculative, as is any physics beyond the standard model, the most conservative approach to explain the existence of the lepton and quark families  likely is to assume that the particles are composite.  In the 1970's the hypothetical constituents of quarks and leptons were given the name ``preons''   \cite{Pati:74}.

The circumstantial evidence for composite leptons and quarks is certainly not conclusively corroborated by direct experimental evidence. The factor $g_\mu$ relates the muon's magnetic moment ${\boldsymbol \mu}$ and  spin ${\boldsymbol s}$ according to the relation,
\begin{equation}\label{eqn:1}
{\boldsymbol \mu}=g_\mu \frac{e\hbar}{2m_\mu c}{\boldsymbol s}\,,\hspace{1.0 cm}g_\mu=2(1+a_\mu)\,.
\end{equation}
The present experimental world average determination for $a_\mu$  \cite{Bennett:04} is
$a_\mu^\mathrm{exp}=116592080(63)\times 10^{-11}\,.$ The primary source of error in the
theoretical calculation of $a_\mu$ arises from the contribution of the strong
force, which cannot be calculated from first principles but,
instead, must be determined from experimental data.  A direct
determination of the contribution from strong interactions uses
experimental data from electron-positron collisions, and the theoretical value calculated from the standard model is \cite{Rafael:09} $a_\mu^{\mbox {SM direct}}=(116591785\pm 51)\times 10^{-11}$.
The theoretical value in  is smaller than the experimental value by 3.6 $\sigma$. The
indirect method   \cite{Davier:09} uses experimental data from hadronic  tau decays,
conservation of the vector current and isospin corrections and yields the value $a_\mu^{\mbox {SM indirect}}=(116591932 \pm 52)\times 10^{-11}$, which is 1.8 $\sigma$ below the experimental value.
The discrepancy suggests that there is physics beyond the standard model.  But in addition to muon substructure,  the discrepancy could also be explained, for example,  by the existence of supersymmetric particles or by composite $W$ and $Z$ bosons.  

If leptons, quarks or both are composite, several characteristics  that any  composite model must possess can immediately be  determined from the mass- and spin-spectra of the leptons and quarks. The masses of the charged leptons and quarks satisfy the inequalities ${\rm m}_{\rm electron}\ll{\rm m}_{\rm muon} \ll {\rm m}_\tau$, ${\rm m}_{\rm up} \ll {\rm m}_{\rm charm} \ll {\rm m}_{\rm top}$, and ${\rm m}_{\rm down} \ll {\rm m}_{\rm strange} \ll {\rm m}_{\rm bottom}$.  Since no structure has been conclusively detected for leptons or quarks, any composite system must be very strongly bound.  This precludes a composite model whereby, for example, the electron,  muon, and tau  are successively  much more massive because one or more of their constituents are successively much more massive. The large differences in the masses of the particles in each family are consistent with strong binding only if the bound system is highly relativistic.

Even though the existence of families is explained by the atomic model of atoms, the quark model for mesons and baryons and, as discussed here, perhaps a preon model for leptons and quarks, each of the composite models is radically different. In the atomic model, aside from small mass defects, the mass of an atom is  the sum of the masses of the constituents.  For mesons and baryons, most of the mass results from the kinetic energy of the quarks so the mass of a meson or baryon is substantially greater than the sum of the masses of the constituent quarks. If leptons, quarks or both are highly relativistic bound states, the mass of each lepton or quark is less - and for the least massive particle in each family, much less - than the sum of the masses of the constituents.

For simplicity the bound system would likely be comprised of two or three constituents.  If the composite system were comprised of a spin-0 boson and a spin-1/2 fermion, then, in the language of non-relativistic physics, if all  states have zero orbital angular momentum, all  would have total angular momentum or spin one half.  For the relativistic equation discussed here, a similar mechanism allows strongly bound states only when the total angular momentum or spin is one half. If the composite system were comprised of three or more constituents, it is difficult  to find a mechanism that would prevent higher-spin bound states. 

Many physicists in the 1970's  studied the possibility that quarks and leptons might be bound states of preons \cite{Marshak:93}. Assuming the existence of only a few preons, the existence of all quarks and leptons could be explained with each being a preon bound state.  In the book {\it The Trouble with Physics} \cite{Smolin:06}, Smolin writes, ``Unfortunately, there were major questions that the preon theories were not able to answer. These have to do with the unknown force that must bind the preons together into the particles that we observe.  The challenge was to keep the observed particles as small as they are while keeping them very light. Because preon theorists couldn't solve this problem, preon models were dead by 1980.''

Leptons interact gravitationally, weakly, and electromagnetically.  Among the three forces, the electromagnetic interaction is the only one that might be able to provide the requisite strong binding. At sufficiently high energies, the electromagnetic interaction must, of course, be replaced by the electroweak interaction, but this added complication is not incorporated in the Bethe-Salpeter equation discussed here.  The similarities of the charged lepton and quark mass spectra suggest that the same mechanism might be responsible for binding in all four families.  Such a  scenario could occur if only one of the two quark constituents interacted strongly so that the pair would not.  A heretofore unknown interaction might, of course, be responsible for the binding, but assuming the existence of such a force would represent a much more speculative approach.

A few nonrelativistic and partially relativistic calculations suggest that electromagnetism might be able to create a bound state with the properties of quarks and leptons. Using the Schr\"odinger equation in two space dimensions,  a charged quanta interacting with a charged magnetic dipole was shown to create bound states such that all low-energy states have zero orbital angular momentum \cite{Mainland:83}.   The Dirac equation was used to show that strongly bound states can occur when a spin-1/2 quanta interacts with a charged magnetic  dipole \cite{Barut:76}, and the Klein-Gordon equation in two space dimensions was used to show that a charged quanta interacting with a charged magnetic dipole can create bound states such that the energy gap increases between successively higher bound states  \cite{Mainland:84}.

But, as previously mentioned,  by about 1980 most physicists had given up on the idea: experimentally there was no conclusive evidence that quarks or leptons were composite, a situation that has not changed significantly in the intervening thirty years, and theoretically no one could find a way to create bound states that  both possessed the correct mass spectra and also were  extremely small.   The Bethe-Salpeter equation \cite{Salpeter:51} provided a theoretical framework for studying relativistic bound states.  Also, by the late 1960's two, two-body, bound-state Bethe-Salpeter equations with unphysical interactions had been completely solved. However, in 1980 a general, numerical technique for solving the two-body, bound-state Bethe-Salpeter equation had not been developed, and the high-speed computers required to solve the bound-state equation when the preons interact via minimal electrodynamics did not exist.  With the development of super computers and a general method for solving the two-body, bound-state Bethe-Salpeter equation \cite{Mainland:05},  the properties of relativistic bound states can now be determined, providing motivation to revisit preon models.

\section{Possible Preon Model of   Leptons and Quarks}

If the charged lepton family, the neutrino family, the negatively charged quark family, and the positively charged quark family are each a bound state of a different combination of a spin-0 and a spin-1/2 preon that interact via minimal electrodynamics, then all four families could be created from just two fermion preons and two  boson preons.  Each of the  two boson preons can bind with each of the  two  fermion preons to create four different bound states or four families as indicated in Table 1.  Also, as indicated in Table 1, the sum of the charges of the two constituent preons must equal the charge of the leptons or quarks in the family that the preons combine to create, yielding four equations for the charges of the four preons.
\begin{table}[ht]
\begin{center}
\caption{Preon charges and constraints that the preon charges must obey}
\begin{tabular}{llll}
Family&Preon&\hspace{0.2 cm}Preon&\hspace{0.4 cm}Constraint\\
&Fermion&\hspace{0.2 cm}Boson&\\
&Charge&\hspace{0.2 cm}Charge &\\
\hline
 Electron&$\hspace{0.3 cm} q_{f1}$&$\hspace{0.4 cm}Q_{S1}$&\hspace{0.4 cm}$q_{f1}+Q_{S1}=-1$\\
Neutrino&$\hspace{0.3 cm} q_{f2}$&$\hspace{0.4 cm}Q_{S1}$&\hspace{0.4 cm}$q_{f2}+Q_{S1}=0$\\
Negative Quarks&$\hspace{0.3 cm} q_{f1}$&$\hspace{0.4 cm}Q_{S2}$&\hspace{0.4 cm}$q_{f1}+Q_{S2}=-\frac{1}{3}$\\
Positive Quarks&$\hspace{0.3 cm} q_{f2}$&$\hspace{0.4 cm}Q_{S2}$&\hspace{0.4 cm}$q_{f2}+Q_{S2}=\frac{2}{3}$\\
\end{tabular}
\label{Table:1}
\end{center}
\end{table}
Only three of the equations are independent. Possible preon charges $\leq 2$ that are multiples of
$\pm\frac{1}{3}$ are as follows:
\begin{table}[h]
\begin{center}
\caption{Possible charges of preon fermions and bosons}
\begin{tabular}{cccc}
\multicolumn{2}{c}{Preon Fermion Charges}&\multicolumn{2}{c}{\hspace{1.5
cm}Preon Boson Charges}\\
\hline
$q_{f1}$&\hspace{1.9 cm}$q_{f2}$&\hspace{1.6 cm}$Q_{S1}$&\hspace{1.4 cm}$Q_{S2}$\\
1&\hspace{1.9 cm}2&\hspace{1.6 cm}-2&\hspace{1.0 cm}$-\frac{4}{3}$\\

$\frac{2}{3}$&\hspace{1.9 cm}$\frac{5}{3}$&\hspace{1.4 cm}
$-\frac{5}{3}$&\hspace{1.0 cm}$-1$\\ 

$\frac{1}{3}$&\hspace{1.9 cm}$\frac{4}{3}$&\hspace{1.5
cm}$-\frac{4}{3}$&\hspace{1.0 cm}$-\frac{2}{3}$\\ 

$\hspace{-0.2 cm}-\frac{4}{3}$&\hspace{1.6 cm} $-\frac{1}{3}$&\hspace{1.8
cm}$\frac{1}{3}$&\hspace{1.3 cm}$1$\\ 

$\hspace{-0.2 cm}-\frac{5}{3}$&\hspace{1.7
cm}$-\frac{2}{3}$&\hspace{1.8
cm}$\frac{2}{3}$&\hspace{1.3 cm}$\frac{4}{3}$\\

$\hspace{-0.2 cm}-2$&\hspace{1.6 cm}$-1$&\hspace{1.8
cm}$1$&\hspace{1.3 cm}$\frac{5}{3}$\\
\end{tabular}
\label{Table:2}
\end{center}
\end{table}

\section{Introduction to the Bethe-Salpeter Equation}

For relativistic, two-body, bound-state systems, a few analytical, Bethe-Salpeter
solutions exist, but only in the limit that the binding energy equals
the sum of the masses of the two constituent particles, implying the the energy 
of the bound state is zero. These solutions are not physical 
because they are calculated in the center-of-mass rest frame. But if
a bound state has zero rest energy, it must travel at the speed of
light.  Therefore, zero-energy solutions are the zero-energy limit of
physical solutions. In this limit the equation is always invariant
under rotations in Minkowski space and so is separable. 

Analytical solutions have been obtained in the zero-energy limit,
for example, for the Wick-Cutkosky Model \cite{Wick:54,Cutkosky:54}
that consists of two scalars with arbitrary, nonzero masses bound by
the exchange of a massless scalar. A variety of exact solutions have
been obtained for two spinors with equal masses that are bound by the
exchange of a massless
scalar \cite{Goldstein:53,Kummer:64,Suttorp:75} or a massless
vector \cite{Nishimura:76}. An exact solution has also been found for
two spinors with unequal, nonzero masses  bound by the
exchange of a massless scalar \cite{Keam:71}.

Even when the energy is zero,  most solutions are numerical.
For example, such solutions have been obtained for two scalars
with unequal, nonzero masses bound by the exchange of a massive
scalar \cite{Brennan:75} and for a scalar and a spinor with arbitrary,
nonzero masses bound by the exchange of a massless
scalar \cite{Mainland:03a,Mainland:03b} or a massless
vector \cite{Mainland:08}. When the energy is nonzero, all
solutions are numerical. 

Only four, two-body, bound-state Bethe-Salpeter equations have been
completely solved.  That is, solved for arbitrary values of energy
and arbitrary values of the masses of the two constituents. (1) The
Wick-Cutkosky Model consists of  two  scalars with arbitrary,
nonzero masses bound by the exchange of a massless
scalar \cite{Wick:54,Cutkosky:54,zurLinden:69a}. (2) The
Scalar-Scalar Model  consists of  two  scalars with arbitrary,
nonzero masses bound by the exchange of a massive
scalar \cite{Schwartz:65,Kaufmann:69,zurLinden:69b,zurLinden:69c}.
(3) The Scalar Electrodynamics Model consists of a scalar and a
fermion with arbitrary, nonzero masses bound by the exchange of a
massless scalar \cite{Mainland:05}. (4) The Fermion-Scalar Model   consists of
a scalar and a fermion with arbitrary, nonzero masses bound by the
exchange of a massive scalar \cite{Mainland:06}.

The interaction of the preons and the electromagnetic field is  described by the renormalizable interaction  Lagrangian,     
\begin{align}\label{eqn:2}
 L_{\rm int} =& : - q_fA^{\mu} \bar{\psi}\gamma_{\mu}\psi+ \imath
Q_s\,A_{\mu}\,[
\phi\,(\partial^{\mu}\phi^\dagger)  
  -(\partial^{\mu}\phi)\,\phi^\dagger]\nonumber\\
  &+Q_s^2A^\mu A_\mu \phi \,
\phi^\dagger:\,.
\end{align}
The proposed preon model of leptons, quarks or both is a bound
state of a spin-1/2 fermion with charge $q_f$ and mass m$_f$ represented
by the field $\psi$ and  a scalar with charge $Q_s$ and mass m$_s$
represented by the field $\phi$. The fermion and scalar fields interact
minimally with the electromagnetic field $A_\mu$.

The Bethe-Salpeter equation will be used to study the
properties of the bound state. Since every Feynman diagram
contributes to the exact equation,  some approximation must be
made. Usually only the effect of the lowest-order diagram shown
below is included.  Because of the structure of the Bethe-Salpeter
equation, contributions from the diagram and iterations of the
diagram, which form a ladder with the photon propagator forming the
rungs, contribute. Thus the approximation is called the ladder
approximation.
\begin{figure}[h]
\centering
\includegraphics[width=80mm]{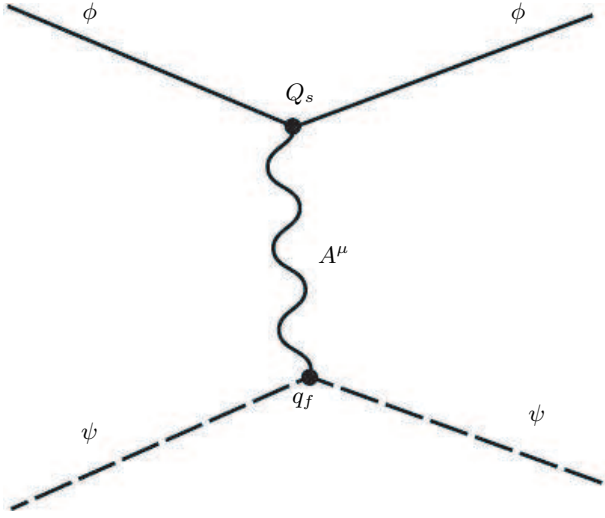}

\vspace{-6.9 cm} $\phi$\hspace{5.5 cm}$\phi$

\vspace{0.7 cm} \hspace{-0.1 cm}$Q_s$

\vspace{1.8 cm} \hspace{0.8 cm}$A^\mu$

\vspace{1.5 cm} $q_f$

\vspace{0.1 cm} \hspace{-5.7 cm}$\psi$

\vspace{-0.6 cm}\hspace{6.2 cm}$\psi$

\vspace{.9  cm}

\caption{Lowest-order  Feynman diagram for a spin-1/2 field
$\psi$ with charge $q_f$ and a scalar field $\phi$ with a
charge $Q_s$ interacting via minimal electrodynamics with the photon
field $A^\mu$.}
\label{fig:1}
\end{figure}

In the ladder approximation  the two-body, bound-state
Bethe-Salpeter equation  describing a spin-0 boson and a spin-1/2
fermion  bound by minimal electrodynamics is 
\begin{align}\label{eqn:3}
 [(p^0&+\xi E) \gamma_0+p^i\gamma_i  -m_f]\nonumber \\
 &\times \{[p^0+(\xi-1)E]^2-{\boldsymbol{p}}^2  -m_s^2\}\chi(p)\nonumber \\
&\hspace{-0.6 cm}=\frac{iq_fQ_s}{(2\pi)^4} \int_{-\infty}^{\infty} \frac{{\rm d}^4q}
{(p-q)^2+i\varepsilon} \{ \gamma^0[p_0+q_0+2(\xi-1)E]\nonumber \\
& + \gamma^i(p_i+q_i)\}\chi(q)\,.
\end{align}
The equation has been written in the center-of-mass rest frame
where the four-momentum $K^\mu$ of the bound state takes the form $K^\mu=(E,0,0,0)$. In the non-relativistic limit  
$\xi=m_f/(m_f+m_s)$.

 If the binding were non-relativistic, the properties of the bound states could be calculated using the Schr\"odinger equation,
\begin{equation}\label{eqn:4}
H(x)\psi(x)=E\psi(x)\,.
\end{equation}
There are a number of notable differences between the bound-state Schr\"odinger  and Bethe-Salpeter equations:  a) Because the energy $E$ appears multiple times in  the Bethe-Salpeter equation (\ref{eqn:3}), a  Hamiltonian does not exist  for the relativistic problem.  b) Because $E$ appears multiple times in the Bethe-Salpeter equation, the equation is solved by specifying the energy  and solving for the coupling constant $q_fQ_s/(4\pi)$ as an eigenvalue.  The process for solving the Schr\"odinger equation  (\ref{eqn:4}) is the reverse:  the coupling constant is specified, and the energy $E$ is solved for as an eigenvalue.   The two procedures yield equivalent information.  (c) There is no action at a distance for the Bethe-Salpeter equation since the interaction is covariant. (d) The Bethe-Salpeter is an integral equation; therefore, boundary conditions are incorporated in the equation. (e) The Bethe-Salpeter equation is separable when energy $E=0$ and is almost always nonseparable  when $E \neq 0$.

 The author has developed a systematic method \cite{Mainland:05} for
solving the two-body, bound-state  Bethe-Salpeter equation for arbitrary values of energy.  However, to show how specific boundary conditions allow for the possibility that strongly bound solutions exist when the coupling constant  $q_fQ_s/(4\pi)$ is on the order of the fine structure constant $\alpha$, here attention is restricted to the strong binding limit $E=0$.

\section{Strongly Bound Solutions}

To solve the Bethe-Salpeter equation in the strong-binding  (zero-energy) limit, the singularity in the kernel of the Bethe-Salpeter equation is removed, and the equation is transformed from Minkowski to Euclidean space by making a Wick rotation \cite{Wick:54}, which is always possible in the ladder approximation. 

In the zero-energy limit the angular dependence of the equation
separates \cite{Sugano:56} when solutions are written as products
of the functions $\Psi^{(+)}_1$ and $\Psi^{(+)}_2$  that are four-component hyperspherical harmonics in four-dimensional, Euclidean space-time
and the  functions $F(|p|)$ and $G(|p|)$ that depend only on
the magnitude of the Euclidean four-momentum $|p|$, 
\begin{eqnarray}\label{eqn:5}
\chi(ip^0,{\bf p})&= F(|p|)
\Psi^{(+)}_1(\cos\,\theta_1,\theta_2,\phi)\nonumber\\
& + G(|p|)
\Psi^{(+)}_2(\cos\,\theta_1,\theta_2,\phi)\,.
\end{eqnarray}
The four-component hyperspherical harmonics   $\Psi^{(+)}_1$ and $\Psi^{(+)}_2$ in (\ref{eqn:5})  are  analogous to the two-component spherical harmonics $\phi^{(\pm)}_{j,m}$ \cite{Bjorken:65} that are used to
separate the Dirac equation with a spherically symmetric potential.
Using hyperspherical harmonics in four-dimensional space as part of the basis functions 
both decreases the number of integrations that must be performed numerically and
prevents the appearance of a logarithmic singularity in the kernel.  The separated, zero-energy Bethe-Salpeter equation is
\alphaeqn
\begin{align}\label{eqn:6a}
&\left[ |p|^2+(1+\Delta)^2 \right]\left[ (1-\Delta)F^{(+)}(|p|)+|p|G^{(+)}(|p|)
\right] \nonumber \\
&= -\frac{q_fQ_s}{(2\pi)^4} |p| \int_0^\infty |q|^3
\Lambda^{(2)}_{k_1+1/2}(|p|,|q|)G^{(+)}(|q|) 	\mathrm{d}|q|\nonumber \\
&- \frac{q_fQ_s}{(2\pi)^4} \int_0^\infty |q|^4
\Lambda^{(2)}_{k_1-1/2}(|p|,|q|)G^{(+)}(|q|) 	\mathrm{d}|q|\,,
\end{align}
\begin{align}\label{eqn:6b}
&\left[ |p|^2+(1+\Delta)^2 \right]\left[ -|p|F^{(+)}(|p|)+(1-\Delta)G^{(+)}(|p|)
\right] \nonumber \\  
&=	 \frac{q_fQ_s}{(2\pi)^4} |p| \int_0^\infty
|q|^3\Lambda^{(2)}_{k_1-1/2}(|p|,|q|)F^{(+)}(|q|) 	\mathrm{d}|q|\nonumber \\
&+\frac{q_fQ_s}{(2\pi)^4}\int_0^\infty
|q|^4\Lambda^{(2)}_{k_1+1/2}(|p|,|q|)F^{(+)}(|q|) 	\mathrm{d}|q|\,.
\end{align}
\reseteqn
In the above equation the index $k_1=1/2, 3/2,\dots$ and
\begin{eqnarray}\label{eqn:7}
	\Lambda^{(2)}_{n}(|p|,|q|) =\left \{
\begin{array}{lcc}
\frac{2\pi^2}{|p||q|}\;\frac{(\frac {|q|}{|p|})
^{n+1}}{n+1}	&	\mathrm{if}& |q| \le |p|\,, \\
\\
\
\frac{2\pi^2}{|p||q|}\;\frac{(\frac{|p|}{|q|})
^{n+1}}{n+1}	&	\mathrm{if}& |q| \ge |p|\,.\\
\end{array}\right.
\end{eqnarray}
Dimensionless variables have been introduced in (6) by defining $m_f \equiv
m(1-\Delta), m_s \equiv m(1+\Delta), p' \equiv p/m$ and $q'\equiv q/m$. 
Then primes have been omitted for all momentum variables since all are dimensionless.
 
 Eq. (6) is solved by expanding the functions $F(|p|)$ and $G(|p|)$ in terms of a finite set of basis functions that (very nearly) obey the boundary conditions that $F(|p|)$ and $G(|p|)$, respectively, obey. 
\begin{align}\label{eqn:8}
\mbox{basis function}= \mbox{convergence function}\times\mbox{spline}
\end{align}
The convergence function depends on  the  magnitude of the 
four-momentum $|p|$ and  obeys the  boundary conditions that the
solution obeys. A general method for  analytically calculating boundary
conditions  for Bethe-Salpeter equations has been
developed \cite{Mainland:03b}. Typically, as is the case for the
Bethe-Salpeter equation being discussed here, the behavior of the
solution at small momenta $|p|$ can be determined exactly; however, at large $|p|$
boundary conditions can sometimes only be determined within ranges,
complicating the calculation. If the solution $F(|p|)$ obeys the boundary
conditions
\begin{equation}\label{eqn:9}
F(|p|)_{\stackrel{\displaystyle\longrightarrow}  {|p| \rightarrow 0}}
|p|^{f_0}\,, \hspace{1.0 cm} 
  F(|p|) _{\stackrel{\displaystyle\longrightarrow}  {|p| \rightarrow \infty}}
|p|^{-f_\infty}\,,\\
 \end{equation}
then the convergence function for  $F(|p|)$  is given by
\begin{equation}\label{eqn:10}
\mbox{convergence function}
=\frac{|p|^{f_0}}{(|p|+{\rm constant})^{f_0+f_\infty}}\,.\\ 
\end{equation}

Cubic splines $B_i(|p|)$ \cite{Boor:78} are defined on
four contiguous intervals, and the value of
$|p|$ at the boundary of each interval is called a ``knot'' that is  denoted by $T_j$. Within
each interval a cubic spline is a cubic polynomial with coefficients
chosen so that the cubic spline and its first two derivatives are
continuous, implying that a cubic spline and its first two
derivatives vanish at the beginning of the first interval and at the
end of the fourth. The single, remaining, unspecified coefficient is
fixed by an overall normalization condition.   The spline
$B_{i}(|p|)$ begins at the knot $T_i$ and terminates at $T_{i+4}$. A
major advantage of using splines, instead of functions defined on the
entire interval $0 \leq |p| \leq\infty$, is that,  by concentrating
knots in the region where the solution is changing most rapidly,
more splines are available where they are most needed to represent a
solution. Additionally, it is much faster to integrate numerically
over splines because they are nonzero only on a finite interval.
\begin{figure}[h]
\centering
\hspace{2.0 cm}
\includegraphics[width=80mm]{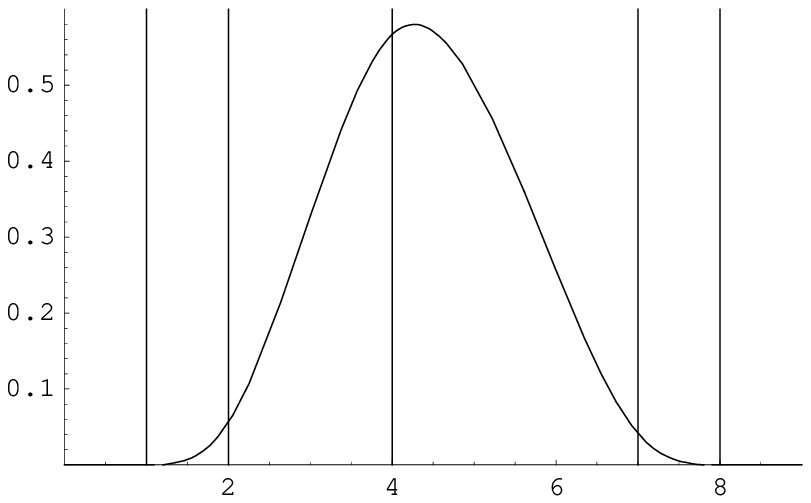}

\vspace{-4.8 cm}

\hspace{1.3 cm} \#1\hspace{0.5 cm} \#2\hspace{1.2 cm} \#3\hspace{1.0
cm} \#4

\vspace{1.8 cm}

\hspace{-7.1 cm} $B_i(|p|)$

\vspace{2.2 cm}

\hspace{1.6 cm} $T_i$\hspace{0.2 cm}$T_{i+1}$\hspace{0.7 cm} $T_{i+2}$\hspace{1.5 cm}$T_{i+3}$\hspace{0.1 cm}$T_{i+4}$\

\vspace{0.1 cm}

\indent$ \hspace{0.9 cm}${$|p|$}

\caption{Cubic spline $B_i(|p|)$ plotted as a function of the magnitude of the
four-momentum $|p|$.}
\label{fig:2}
\end{figure}

Because the convergence function obeys the boundary conditions exactly  at large and small $|p|$, and because the splines also depend on $|p|$, the basis functions very nearly obey the boundary conditions, but they do not obey the boundary conditions exactly.  However, linear combinations of these basis functions obey the boundary conditions exactly and almost always yield convergent series expansions for solutions.

The zero-energy equation is discretized  and then solved by converting
it into a generalized matrix eigenvalue equation using the
Rayleigh-Ritz-Galerkin variational method \cite{Delves:74,Atkinson:76}. 

Typically  strongly bound solutions of the Bethe-Salpeter equation exist only when the coupling constant $q_fQ_s/(4\pi)$ is on the order of or greater than unity.  If  solutions exist for  $q_fQ_s/(4\pi) \sim \alpha$, then at least one integral  in (6) must be very large. Two sets of boundary conditions \cite{Mainland:08} could yield one or more large integrals, and in both cases  the integrals could be large when the coupling constant is small only  if the  spin of the bound state is one half.  If leptons, quarks or both are composite, this mechanism would explain why leptons, quarks or both all have spin one half.  Here attention is restricted to the following boundary conditions:
\alphaeqn
\begin{align}\label{eqn:11a}
&F(|p|)\xrightarrow[|p|
\rightarrow
\infty]{} F_\infty \;|p|^{-(k_1+\frac{7}{2})}\,,\\ 
\label{eqn:11b} 
&G(|p|)\xrightarrow[|p| \rightarrow
\infty]{} G_\infty \;|p|^{-(k_1+\frac{5}{2}+\epsilon_g)}\,.
\end{align}
\reseteqn
In (11), $k_1=j, j+1,\dots$, and $F_\infty$ and $G_\infty$ are constants. The parameter $\epsilon_g$ satisfies the conditions $0 < \epsilon_g < 2$ if $k_1=1/2$, and  $0 \leq \epsilon_g <2$ if $k_1\geq 3/2$.

For solutions satisfying the boundary conditions (11), in the limit of large $|p|$, the Bethe-Salpeter equation (\ref{eqn:6a}) decouples from (\ref{eqn:6b}) and takes the form
\begin{align}\label{eqn:12}
&G_\infty|p|^{-k_1+1/2-\epsilon_g} \xrightarrow[|p|
\rightarrow
\infty]{}  -\frac{q_fQ_s}{4\pi}\frac{1}{2\pi}\times\nonumber\\
&\left [\frac{2(k_1+1)}{(k_1+\frac{1}{2})(k_1+\frac{3}{2})}|p|^{-(k_1+3/2)}\int_0^{|p|}{\rm d}|q||q|^{k_1+7/2}G(|q|)\right .\nonumber\\
&+\frac{|p|^{k_1+3/2}}{k_1+\frac{3}{2}}G_\infty \int_{|p|}^\infty {\rm d}|q||q|^{-(2k_1+2+\epsilon_g)} \nonumber\\
&\left .+\frac{|p|^{k_1-1/2}}{k_1+\frac{1}{2}}G_\infty \int_{|p|}^\infty {\rm d}|q||q|^{-(2k_1+\epsilon_g)} \right ]\,.
\end{align}
Only the final integral in (\ref{eqn:12}) can be  large and then iff $k_1=1/2$ and $\epsilon_g \ll 1$. Since $k_1=j, j+1, \dots$, and $k_1=1/2$ for all strongly bound states with small coupling constants, all such states have total angular momentum  or spin one half. 

Using expressions for the  integrals  in (\ref{eqn:12}) that are valid in the limit of large $|p|$ \cite{Mainland:03b}, the following expression is obtained for the coupling constant when $k_1=1/2$:
\begin{equation}\label{eqn:13}
\frac{q_fQ_s}{4\pi}=-\frac{\pi}{2}\frac{\epsilon_g(2-\epsilon_g)(2+\epsilon_g)}{(1+\epsilon_g)}\,.
\end{equation}
For small values of $\epsilon_g$, the coupling constant $q_fQ_s/(4\pi)$ is small.

 It is particularly difficult to obtain solutions when a fermion and scalar interact via minimal electrodynamics instead of through the exchange of a massless or massive scalar. As can be seen from the dependence of the coupling constant on $\epsilon_g$, the solutions go to zero at large momenta sufficiently slowly that the behavior of solutions at large momenta has a very significant effect on the coupling constants.\\

\section{Summary}

Three major points are made in the talk: a) It might be possible to describe leptons, quarks
or both as highly relativistic bound states of a spin-0 and spin-1/2
constituent bound by minimal electrodynamics. b) Strongly bound states with coupling
constants on the order of the electromagnetic fine structure constant $\alpha$
are allowed by the boundary conditions and likely exist.  c) All such strongly bound states would have
spin one half, as do the leptons and quarks.

\begin{acknowledgments}
This work was supported by a grant from The Ohio Super Computer Center
\end{acknowledgments}

\bigskip 
\bibliography{dpf2009_GBMainland}




\end{document}